\documentclass[aps,prl,twocolumn,showpacs,superscriptaddress,groupedaddress]{revtex4}  
\usepackage{graphicx}  
\usepackage{dcolumn}   
\usepackage{bm}        
\usepackage{amssymb}   
\usepackage{amsmath}
\usepackage[caption=false]{subfig}
\usepackage{natbib}

\begin{document}


\title{Programming complex shapes in thin nematic elastomer and glass sheets}
\author{Paul Plucinsky} 
\affiliation{Engineering and Applied Science, California Institute of Technology, Pasadena, California 91125, USA}
\author{Marius Lemm}
\affiliation{Department of Mathematics,  California Institute of Technology, Pasadena, California 91125, USA}
\author{Kaushik Bhattacharya$^{1}$}
\date{\today}

\begin{abstract}
Nematic  elastomers and glasses are solids that display  spontaneous distortion under external stimuli.   Recent advances in the synthesis of sheets with controlled heterogeneities have enabled their actuation  into non-trivial shapes with unprecedented energy density.  Thus, these have emerged as powerful candidates for soft actuators.  To further this potential, we introduce the key metric constraint which governs shape-changing actuation in these sheets.  We then highlight the richness of shapes amenable to this constraint through two broad classes of examples which we term {\it nonisometric origami} and {\it lifted surfaces}. Finally, we comment on the derivation of the metric constraint, which arises from energy minimization in the interplay of stretching, bending and heterogeneity in these sheets. 
\end{abstract}

\maketitle

Nematic liquid crystal elastomers (LCEs) and glasses (LCGs) are rubbery solids which couple the entropic elasticity of a cross-linked polymer network and the nematic anisotropy of liquid crystals pendent to this network.  The result is a solid displaying many unique mechanical phenomena including large spontaneous distortion due to heating or cooling \cite{wetal_science_15,wb_natmat_15,mw_phystoday_16,detal_angchemie_12,tt_epj_01,wt_book}.

Modes et al.\ \cite{mbw_prsa_10} predicted  that if one could program azimuthal or radial heterogeneity in the anisotropy of a thin nematic glass sheet, then a uniform temperature change would actuate a conical or saddle-like three-dimensional shape.  Such heterogeneity was synthesized by de Haan et al. \cite{detal_angchemie_12} for thin nematic LCG sheets, and thermal actuation of these samples was consistent with the theoretical prediction.  However, the response was muted in light of the small strains and high stiffness of glasses.   
Recently, Ware et al. \cite{wetal_science_15} used novel synthesis techniques in soft and lightly cross-linked LCEs to dramatically realize the predictions of Modes et al.\ \cite{mbw_prsa_10}.  Remarkably, these soft elastomers actuate with volumetric work capacities of 3.6 kJ/m$^3$, comparable to some of the best actuator materials.
Since then, a range of Gaussian curvature has been explored theoretically and achieved experimentally \cite{m_pre_15,mwww_prsa_16,mw_phystoday_16}, and using a metric formalism \cite{esk_sm_13} an explicit recipe for constructing surfaces of revolution from nematic sheets was provided by  Aharoni et al.\ \cite{ask_prl_14}.  

In this letter, we further explain the richness of the shape-changing deformations of LCEs and LCGs and how this can be exploited to make the material act as a machine \cite{bj_science_05}.
The foundation of our work is the metric constraint governing actuation in these sheets (equation (\ref{eq:metric2D}) below).  
We start from an established theory of LCEs by Bladon et al.\ \cite{btw_pre_93,wt_book} and show that designs and deformations that satisfy (\ref{eq:metric2D}) arise naturally from energy minimization.  We sketch the derivation at the end of the letter and refer to the companion paper \cite{plb_inprep} for details.

This metric constraint is a generalization of the metric constraint underlying \cite{ask_prl_14} with two novel features which dramatically expand the design landscape for shape-changing deformation in these sheets.   First, smoothness is not a requirement here. With this, we explore {\it nonisometric origami} where heterogeneity is programmed in a piecewise constant pattern so that thermal actuation leads to complex folding patterns.  
Second, the constraint is amenable to three dimensional programming.  With this, we explore {\it lifted surfaces} where heterogeneity is programmed so that thermal actuation leads to a prescribed surface of arbitrary complexity as long as it is smooth and has limited slope.  


To introduce the key metric constraint, we focus on LCEs noting that the results may be adapted to LCGs with  minor modifications.  Let the unit vector $n_0 \in \mathbb{S}^2$ denote the nematic director or the direction of anisotropy, and let $\bar{r}(T) \ge 1$ be a temperature dependent parameter which captures the stretch along the director and contraction transversely.  This parameter is assumed to be monotonically decreasing for temperature below the isotropic-nematic transition temperature and equal to $1$ in the isotropic regime.  Thus, for a nematic-genesis LCE formed at temperature $T_0$ and subjected to a new temperature $T_f$, a spontaneous distortion with stretch $\ell_{n_0}^{1/2}$ is the preferred state, where
\begin{align}\label{eq:metric3D}
\ell_{n_0} := r^{-1/3}(I_{3\times3} + (r - 1)n_0 \otimes n_0)
\end{align}
is the step-length tensor \cite{wt_book} and $r = \bar{r}(T_f)/\bar{r}(T_0)$, so that $r>1$ for cooling and $r \in (0,1)$ for heating.  

For actuation, we consider a thin sheet of thickness $h$ occupying an initially undeformed flat three dimensional region $\Omega_h := \omega \times (-h/2,h/2) \subset \mathbb{R}^3$ where $\omega \subset \mathbb{R}^2$ denotes the two dimensional midplane of the sheet.   In the synthesis of LCEs sheets (e.g \cite{wetal_science_15}), typically $h \sim 10 \mu m$ whereas the lateral dimensions of the sheet are much larger, typically $\sim cm$.  Hence, we assume $h \ll 1$ and the characteristic lengthscale of $\omega$ is $O(1)$ in non-dimensional units.  Let $x := (x_1,x_2,x_3)$ denote the position on $\Omega_h$ in a Cartesian frame with $\{e_1,e_2,e_3\}$ denoting the basis and $e_3$ pointing normal to $\omega$. We will identify a point $x' := (x_1, x_2) \in \omega$ with $(x_1,x_2,0) \in \Omega_h$.  By a program or design, we mean the prescription of a non-uniform director field on the sheet $n_0: \Omega_h \to \mathbb{S}^2$.  In this letter, we only consider directors that are uniform through the thickness, i.e., $n_0=n_0(x')$.  When the sheet is heated or cooled, non-uniform spontaneous distortion forces a possible out-of-plane deformation of the sheet.  If the sheet is thin enough (we return to this later), it suffices to study the deformation of the midplane,  $y: \omega \to {\mathbb R}^3$.  In particular, we are interested in midplane deformations which are stress-free. These are characterized by the metric constraint
\begin{align}\label{eq:metric2D}
(\nabla ' y)^T \nabla' y&= r^{-1/3}(I_{2\times2} + (r-1) n_0' \otimes n_0' ) =: \ell_{n_0}'
\end{align}
almost everywhere on $\omega$.  Here, $\nabla'$ is the planar gradient (i.e., with respect to $x'$) so that $\nabla' y$ is a $3\times2$ matrix, $n_0' := (n_0\cdot e_1, n_0 \cdot e_2)$ is the projection of $n_0$ onto the plane $\omega$ and $\ell_{n_0}'$ is the $2 \times 2$ submatrix of $\ell_{n_0}$ associated to this projection.  Note that since $n_0'$ is a projection, it need not be a unit vector.

\begin{figure}
\centering
\includegraphics[width=0.4\textwidth]{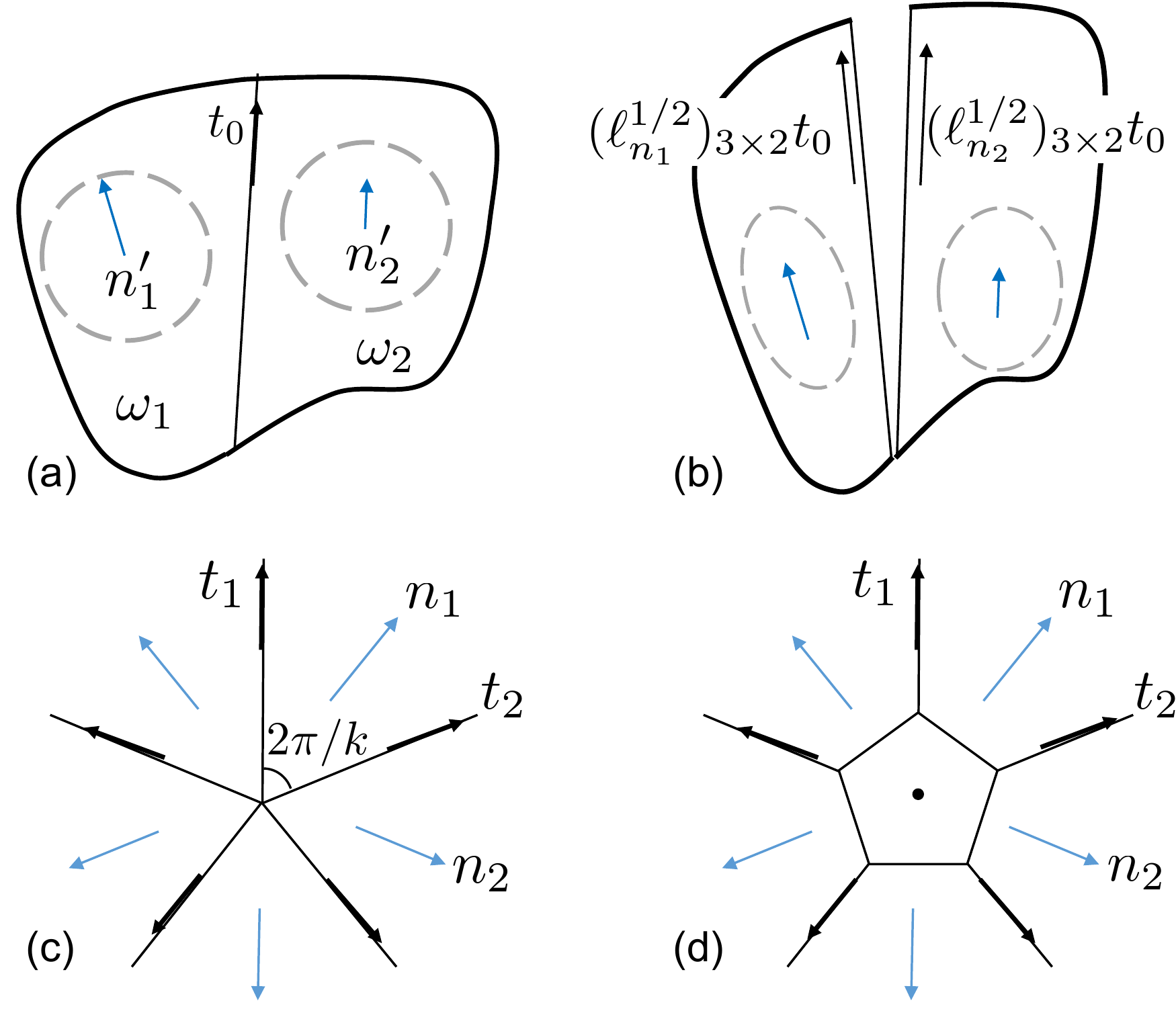}
\caption{Interfaces and junctions on cooling.  (a,b) If each half of the sheet (a) is allowed to independently deform spontaneously, it has the shape in (b) where $(\ell_{n_i}^{1/2})_{3\times2} := r^{-1/6}(I_{3\times2} +(\sqrt{r}-1) n_i \otimes n_i')$.  The interface can be unbroken by rotating one side relative to the other if and only if (\ref{eq:Compat}) holds.  (c) Symmetric junction.  (d) Truncated junction. 
\label{fig:comp}}
\end{figure}

As already intimated, the metric constraint (\ref{eq:metric2D})  generalizes the constraint of Aharoni et al. \cite{ask_prl_14} in two directions; by relaxing the smoothness requirement and by extending the constraint to three dimensional programming.  Indeed, for the former, the metric constraint (\ref{eq:metric2D}) need only hold almost everywhere (i.e., except on sets of zero measure in $\mathbb{R}^2$), and this allows for piecewise constant director designs.   For the latter, (\ref{eq:metric2D}) allows for three dimensional programming while reducing to the constraint of \cite{ask_prl_14} in the case of a planar director.   To see this, if $n_0$ is planar, then $n_0 \equiv n_0'$ and we can write $n_0 \cdot e_1 = \cos(\theta)$ and $n_0 \cdot e_2 = \sin(\theta)$.  It follows that $(\nabla' y)^T \nabla' y = \ell_{n_0}' = R(\theta) \text{diag }(r^{2/3}, r^{-1/3}) R(\theta)^T = g$ for $R(\theta)$ a rotation of $\theta$ about the normal to the initially flat sheet as required by \cite{ask_prl_14}.  

We turn now to examples which highlight the richness of designable surfaces satisfying the metric constraint (\ref{eq:metric2D}).  In addition, these examples serve to motivate the appropriate compatibility conditions consistent with (\ref{eq:metric2D}) for a general class of smooth and non-smooth designable surfaces.  Finally, an important attribute of these designable surfaces is that the actuation is extremely robust since the entire sheet participates in the deformation. This  was observed experimentally in \cite{wetal_science_15}, and it is in marked contrast to other  attempts at foldable structures and origami where the actuation is limited to folds \cite{rhsg_mrl_16},\cite{fwbrvwj_sm_2015}, or bendable structures where through thickness non-uniformity results in complex shape but with little ability to carry load \cite{wmc_prsa_10}.

We begin with {\it nonisometric origami} where the director is programmed in a piecewise constant pattern (also see \cite{mw_phystoday_16, mw_pre_11}).  To start, assume the sheet $\omega$ is the union of two regions $\omega_1$ and $\omega_2$ separated by a straight interface assigned a tangent vector $t_0 \in \mathbb{S}^1$.  Suppose we program this sheet with the director $n_1$ in $\omega_1$ and $n_2$ in $\omega_2$.  Then, it is possible to satisfy (\ref{eq:metric2D}) via a continuous piecewise affine deformation $y$ on all of $\omega$ if and only if
\begin{align}\label{eq:Compat}
|n_1' \cdot t_0| = |n_2' \cdot t_0|
\end{align}
where again $n_i'$ denotes the projection of $n_i$ onto $\omega$.  This is the consequence of a geometric argument for constructing continuous piecewise affine deformations with prescribed metric or stretch tensor provided in Fig. \ref{fig:comp}a-b, an argument that has been applied previously in the study of active martensitic sheets \cite{bj_jmps_99,detal_jap_04}.

\begin{figure*}
\centering
\subfloat[Box\label{sfig:Box}]{%
  \includegraphics[width=0.30\textwidth]{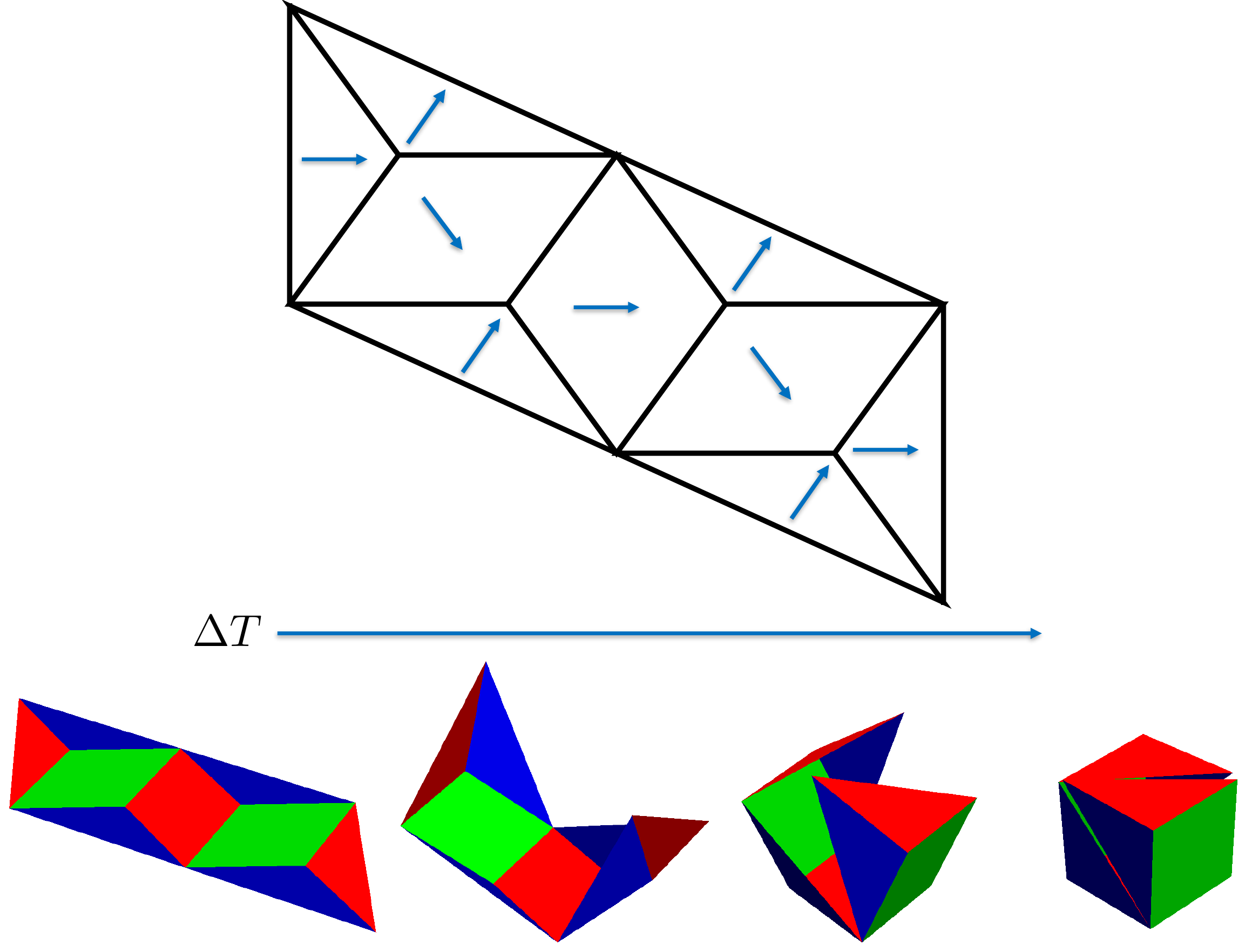}%
}
\hspace{0.5cm}
\subfloat[Rhombic Triacontahedron \label{sfig:Sphere}]{%
  \includegraphics[width=0.30\textwidth]{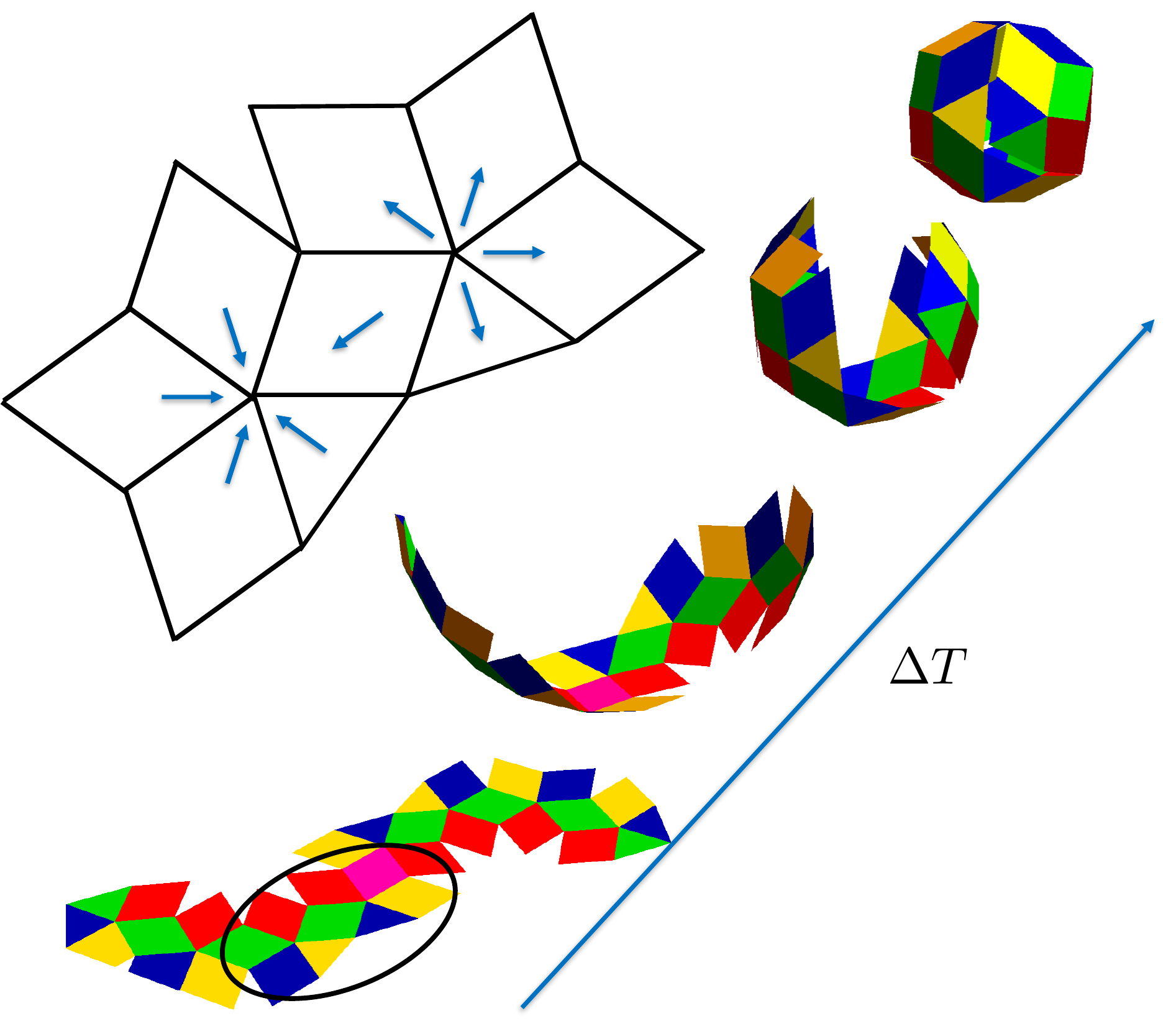}%
}
\hspace{0.5cm}
\subfloat[Devil's Golfcourse \label{sfig:Devil}]{%
  \includegraphics[width=0.30\textwidth]{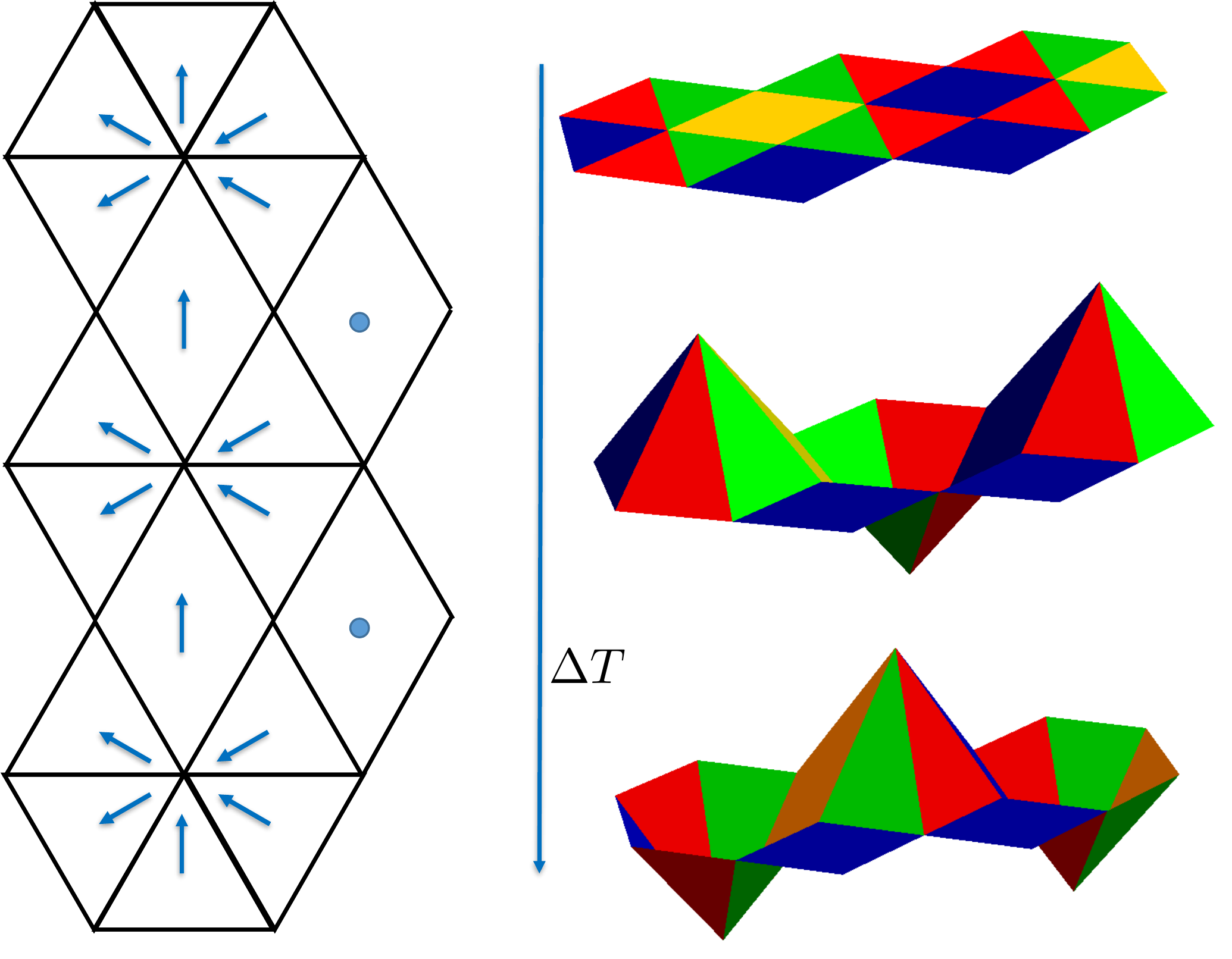}%
}
\caption{Selected examples of nonisometric origami: The line diagrams show the design with the arrows representing the constant director prescribed in each region. The color images show the deformed shape upon cooling.  We note that the designs in (a) and (b)  are compositions of a number of symmetric junctions shown in Fig. \ref{fig:comp}c ($k=3$ in (a) and $k=5$ in (b)).  The design in (c) is a composition of symmetric junctions with $k=6$ and a generalization of the truncated junctions shown in Fig. \ref{fig:comp}d.  This design can be continued periodically.}
\label{fig:Oragami}
\end{figure*}

Now consider a sheet of $k$ sectors $\omega_i, i=1, \dots k$, with the interfaces $t_i$ meeting at a junction and with the sheet programmed with the director $n_i$ in the sector $\omega_i$.  While the condition (\ref{eq:Compat}) is necessary at each interface, it is not sufficient to satisfy (\ref{eq:metric2D}) via a continuous piecewise affine deformation.  One needs an additional global condition to ensure that all the rotations match up as one goes around the junction.  This is extremely rich in general: for example, the case of three sectors with fixed distinct planar directors $n_i \equiv n_i'$ for $i = 1,2,3$ can have up to 32 non-trivial compatible junctions for various $r$ and $n_i$  \cite{p_thesis_16}.  Here though, we focus on a simple case of a junction with all sectors spanning the same angle and with the director programmed to be planar.  In this case, it is possible to satisfy (\ref{eq:metric2D}) via a continuous piecewise affine deformation on cooling (respectively heating) if the director $n_i$ is programmed to bisect the angle between $t_i$ and $t_{i+1}$ (respectively is normal to the bisector) as shown in Fig. \ref{fig:comp}c.  Indeed, on cooling, the angle to each sector reduces, but all the sectors can be brought into contact by rotating them out of plane to form a $k-$sided pyramid.  Note that there is a symmetry here and one can form two possible pyramids (going up or down).  However, one can break this symmetry in practice by adding a small inhomogeneity though the thickness to bias bending in one direction.  One can form a truncated pyramid by replacing the junction with a regular $k-$sided polygon as shown in Fig. \ref{fig:comp}d; each sector is programmed with a planar director as before while the central polygon is programmed with the director to be fully out of plane.

Importantly, it is possible to arrange a number of these junctions and truncated junctions to form complex shapes as we explain with three examples.  First, we can put together a number of three-sided junctions to form a cube as in Fig. \ref{sfig:Box}  (also see \cite{mw_phystoday_16}).  As the temperature decreases and thus $r$ increases each junction becomes a pyramid and eventually becomes the corner of a cube at $r=3$.  Our next example in Fig. \ref{sfig:Sphere} shows a rhombic triacontahedron.  This design is formed by repeating the pattern shown.  Finally, we form a Devil's Golfcourse using the design shown in Fig. \ref{sfig:Devil}.  Since this design is periodic, it can be extended ad infinitum.  We emphasize that these are but a small number of exemplars and many generalizations are possible. (For instance, one can patch an even number of regular polygons into a ring and follow a construction similar to the Devil's Golfcourse to obtain an azimuthally periodic compatible shape.)


We now consider our second class of examples, that of {\it lifted surfaces}.  We look for designs where cooling the sheet leads to a surface that can be described by the graph of a function $\varphi$.  We show that this is possible if function $\varphi$ is smooth enough (in the Sobolev space $W^{2,\infty}$) and satisfies the constraint 
\begin{align}\label{eq:boundVarphi}
\| \nabla' \varphi\|^2_{L^{\infty}} < \lambda_{r} := r - 1
\end{align}
on its domain.  Specifically, we show that we can achieve this shape with the director programmed as follows
\begin{align}\label{eq:designRef}
n_0(x') = \frac{1}{\lambda_r^{1/2}}  \left(\begin{array}{c} \partial_1 \varphi( r^{-1/6} x' )  \\
\partial_2 \varphi(r^{-1/6}x') \\
(\lambda_r - |\nabla' \varphi(r^{-1/6}x')|^2)^{1/2} \end{array}\right) 
\end{align}
and through a deformation $y$ that consist of a uniform contraction followed by a lifting: 
\begin{align}\label{eq:ansatz}
y(x_1,x_2) = r^{-1/6} (x_1 e_1 + x_2e_2) + \varphi(r^{-1/6}x') e_3 .
\end{align}

\begin{figure*}
\centering
\hspace*{0cm}
\subfloat[Caltech\label{sfig:Caltech}]{%
  \includegraphics[width=7.2cm]{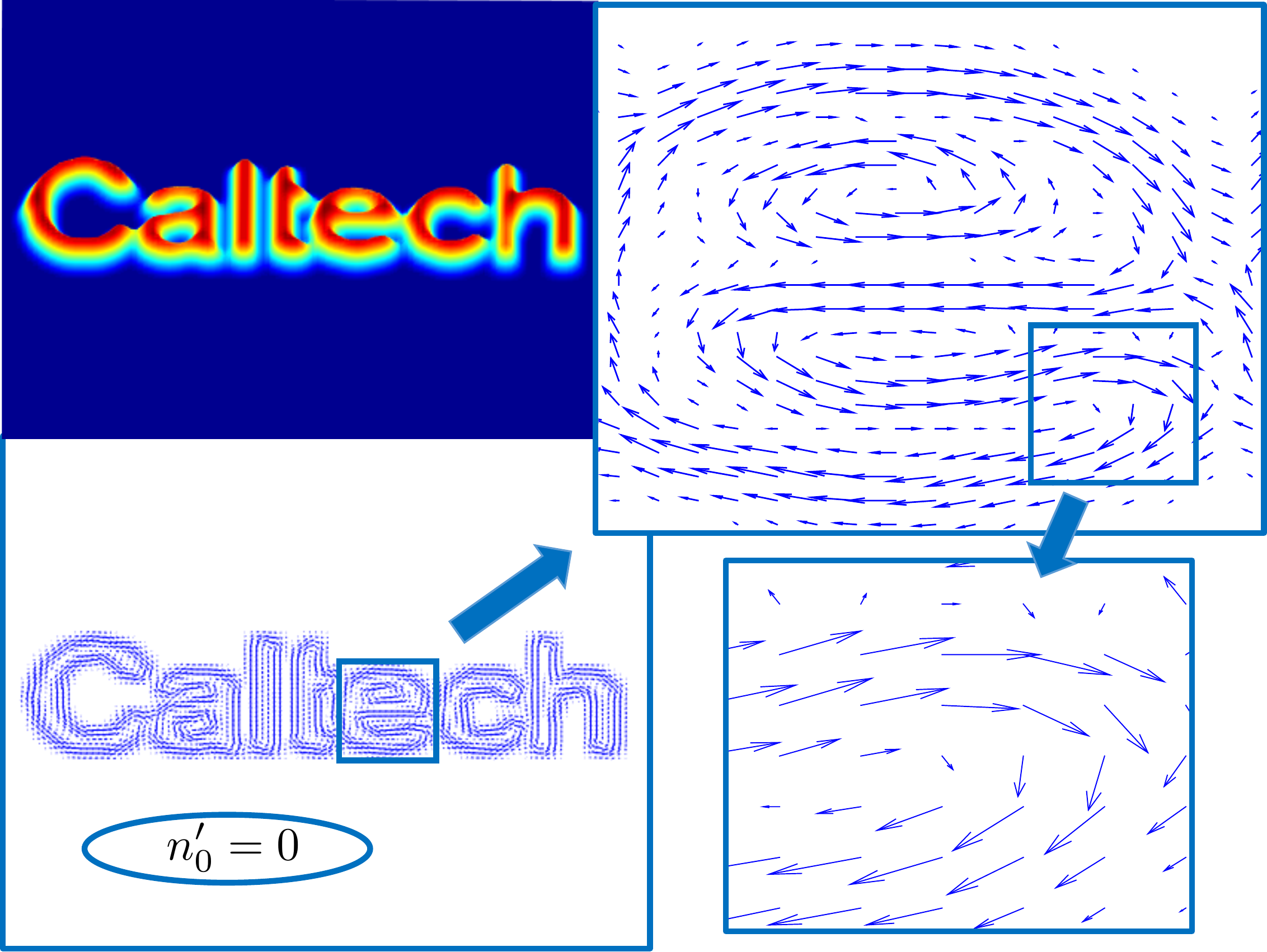}%
}
\hspace*{1.0cm}
\subfloat[Eiffel Tower\label{sfig:Eiffel}]{%
  \includegraphics[width=7.7cm]{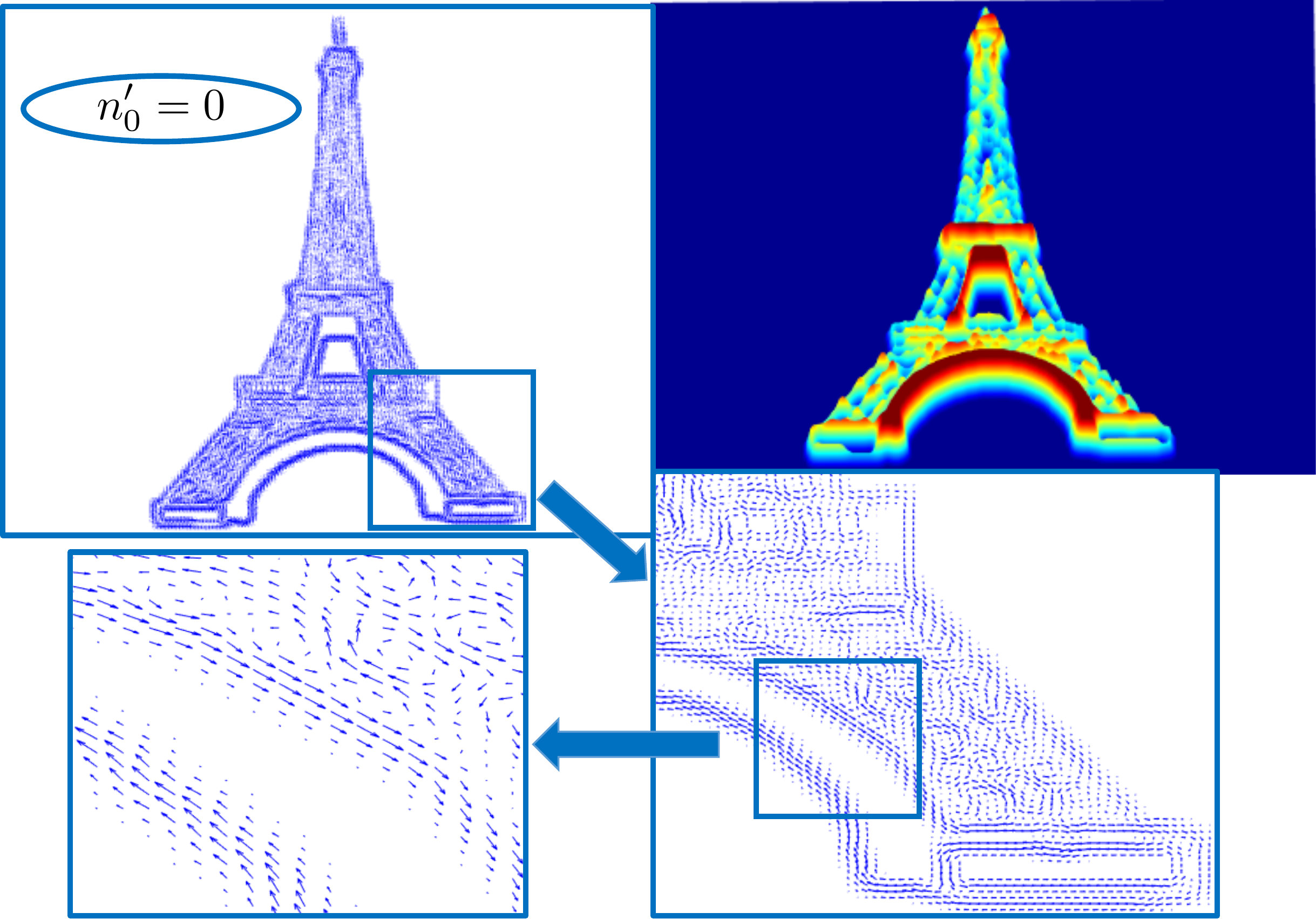}%
}
\caption{The deformed shape and designs for lifted surfaces.  The vector plots show the director orientation in the design.  The amplitude of each vector denotes the planar component $n_0'$ of the director.  The color images show the topographic map of the sheet after deformation with the colors representing height (hot colors are high).  The designs are generated from equation (\ref{eq:designRef}) by taking $\varphi$ to be a smoothened and rescaled greyscale of the desired image.
   }
\label{fig:ComplexShape}
\end{figure*}

Before we prove that this ansatz satisfies the metric constraint (\ref{eq:metric2D}), we note that one can create a large number of shapes using such an approach.  Since $r$ can be significantly different from 1 in LCEs, one can form shapes with significant displacement like spherical caps and sinusoidally rough surfaces.   Fig. \ref{fig:ComplexShape} shows two additional examples with complex surface relief.   These are but a small sample of the designs amenable to this framework.   Indeed, given any arbitrary greyscale image $\mathcal{G}$, we can program a nematic sheet so that the surface of the sheet upon cooling corresponds to this image.  We do this by smearing $\mathcal{G}$ (for instance by mollification or by averaging over a small square twice) and taking this as $\varphi$.

The fact that the lifted surface ansatz satisfies   (\ref{eq:metric2D}) can be verified directly.  However, to motivate the ansatz we now rewrite (\ref{eq:metric2D}) in an equivalent form which points to a concrete design scheme.  Heuristically, we turn the statement around by first identifying the set of deformation gradients consistent with (\ref{eq:metric2D}) for any director and then identifying the director  associated with the deformation gradient.  We conclude that the metric constraint (\ref{eq:metric2D}) holds if and only if 
\begin{align}\label{eq:rewriteMet}
\nabla' y(x') \in \mathcal{D}_r, \;\; n_0(x') \in \mathcal{N}^{\;r}_{\nabla' y(x')} 
\end{align}
for almost every $x' \in \omega$ and
\begin{align} \label{eq:Dnemr}
&\mathcal{D}_{r} := \big\{F' \in \mathbb{R}^{3\times2}: 
 |F'|^2 \le r^{-1/3}+r^{2/3},  \nonumber \\
&r^{-1/3} \leq  |F'e_\alpha|^2 \leq r^{2/3} \;\;  \alpha = 1,2, \\
 &(F' e_1 \cdot F' e_2)^2 = (|F'e_1|^2 - r^{-1/3})(|F'e_2|^2 - r^{-1/3}) \big\}, \nonumber 
 \end{align}
 and
\begin{align}\label{eq:DF}
&\mathcal{N}^{\;r}_{F'} := \big\{ n_0 \in {\mathbb S}^2: (n_0 \cdot e_\alpha)^2 = \frac{|F' e_\alpha|^2 - r^{-1/3}}{r^{2/3} -r^{-1/3}}, \alpha = 1,2,  \nonumber \\
&\text{sign}((n_0\cdot e_1)(n_0\cdot e_2)) = \text{sign}(F' e_1 \cdot F' e_2) \big\}
\end{align}
when $r>1$ (the inequalities for $\mathcal{D}_r$ and the sign in ({\ref{eq:DF}) are switched when $r<1$).
With this description, we seek the restrictions on the class of deformations of the the form (\ref{eq:ansatz}) that satisfy the first condition of (\ref{eq:rewriteMet}).  We find (\ref{eq:boundVarphi}) is sufficient.  We then seek the restrictions on the director $n_0$ that satisfy the second condition of (\ref{eq:rewriteMet}) for this deformation, and this yields the formula (\ref{eq:designRef}).

Naturally, given this analysis, it would be appealing to have a characterization of the geometry of surfaces described by deformations which satisfy (\ref{eq:rewriteMet}) without the ansatz (\ref{eq:ansatz}).   We would then be able to characterize all possible shapes that could be thermally actuated from programing nematic anisotropy into a thin sheet.  Unfortunately, such a broad characterization remains open.  

Finally, we turn to the derivation of the metric constraint (\ref{eq:metric2D}). Our starting point is the well-accepted theory of Bladon et al. \cite{btw_pre_93}.  A LCE formed at temperature $T_0$ with initial director $n_0 \in \mathbb{S}^2$, then subjected to a three dimensional deformation gradient $F \in \mathbb{R}^{3\times3}$ and current director $n\in \mathbb{S}^2$ at temperature $T_f$ has a free energy density given by the non-negative quantity
\begin{align}
\mathcal{F} (F,n,n_0) := \frac{\mu}{2} \left(\text{Tr} (F^T (\ell_n^f)^{-1} F \ell_{n_0}^0 ) - 3\right)
\end{align}
where $\ell_n^f$ and $\ell_{n_0}^0$ are the step-length tensor (\ref{eq:metric3D}) with $r$ replaced by $\bar{r}(T_f)$ and $\bar{r}(T_0)$ respectively.  The incompressibility of elastomers, i.e., $\det F = 1$, is assumed here.  Now, given a thin sheet $\Omega_h$ of thickness $h$ and a design $n_0$, we suppose a three dimensional deformation $y^h: \Omega_h \to {\mathbb R}^3$ of this sheet has a strain energy given by
\begin{align}\label{eq:Ih}
{\mathcal I}_{n_0}^h (y^h) := \int_{\Omega_h} \mathcal{F} (\nabla y^h,\frac{\nabla y^h n_0}{|\nabla y^h n_0|},n_0)dx.
\end{align}
Here, $\nabla$ is the three dimensional gradient as $y^h$ depends on $x = (x',x_3)$, and we introduce a kinematic ansatz on the current director $n^h \colon \Omega_h \rightarrow \mathbb{S}^2$ (the middle argument in $\mathcal{F}$) justifiable for low energy deformations \cite{mbw_prsa_10}.   

To arrive at the metric constraint (\ref{eq:metric2D}), we first observe that due to incompressibility and the kinematic ansatz, $\mathcal{I}_{n_0}^h(y^h)$ is minimized and equal to $0$ if and only if 
\begin{align}\label{eq:3DConstraint}
(\nabla y^h)^T \nabla y^h = \ell_{n_0} \quad \text{ almost everywhere on } \Omega_h
\end{align}
for the three dimensional step-length tensor $\ell_{n_0}$ in (\ref{eq:metric3D}).  However, this equation is not useful for design since it highly restricts the nature of heterogeneity for said program $n_0 \colon \Omega_h \rightarrow \mathbb{S}^2$ (see for instance the discussions in \cite{esk_sm_13}).  Fortunately, it can be relaxed considerably by taking advantage of the thinness of nematic sheets.  In fact, if the thickness $h$ is sufficiently small, it suffices to ignore the constraints associated with the out-of-plane deformation gradient $\partial_3 y^h$ entirely, and focus solely on the satisfying the constraint at the midplane $\omega$.  In doing this, we derive (\ref{eq:metric2D}) from (\ref{eq:3DConstraint}).   

To justify this, we note that generic deformations have energy $\mathcal{I}^h_{n_0}(y^h) = O(h)$.  Thus for designable actuation, we appeal to energy minimization by characterizing deformations for which $\mathcal{I}^h_{n_0}(y^h) \ll O(h)$.  We show that for midplane fields $(y,n_0)$ satisfying (\ref{eq:metric2D}) and for sheets of sufficiently small thickness $h$, we can construct low energy global deformations $y^h$ satisfying $y^h(x',0) \approx y(x')$ in the appropriate Sobolev norm.  For nonisometric origami our constructions satisfy $\mathcal{I}^h_{n_0}(y^h) \leq O(h^2)$, and for sufficiently smooth surfaces such as lifted surfaces our constructions satisfy $\mathcal{I}^h_{n_0}(y^h) \leq O(h^3)$.  The techniques employed here are akin to those of \cite{cd_cvpd_09} for incompressibility and \cite{cm_arma_08} for nonisometric origami.  



Conversely, it is natural to wonder whether (\ref{eq:metric2D}) is an essential feature of low energy configurations.  We show that this is true if we augment the entropic elasticity $\mathcal{I}^h_{n_0}$  studied here with an appropriate version of the Frank elasticity which is natural to nematics.  Specifically, if $\mathcal{\tilde{I}}^h_{n_0}$ is the sum of $\mathcal{I}^h_{n_0}$ and an additional term approximating Frank elasticity, then we can use geometric rigidity \cite{fjm_cpam_02} to show that all bending configurations (i.e., $\mathcal{\tilde{I}}_{n_0}^h(y^h) \leq O(h^3)$) are characterized by sufficiently smooth midplane fields satisfying  (\ref{eq:metric2D}).  

Together these results show two important properties of nematic sheets:  first, that the constraint (\ref{eq:metric2D}) ensures low energy deformations for the sheet ($O(h^2)$ or smaller), and second, that deviation from this constraint results in significant energy ($\gg O(h^3)$ and likely $O(h)$).  This means that the shapes consistent with this constraint are both good  candidates for actuation and robust to added forcing (as observed by Ware et al. \cite{wetal_science_15}).

In closing, we recall that a key ingredient to the design of lifted surfaces is the ability to program the director three dimensionally.  To our knowledge, experimental studies on LCE and LCG sheets have examined planar inscription of the director field \cite{wetal_science_15},\cite{detal_angchemie_12}, but not the case of a fully three dimensional director field.  We hope that promising designs such as lifted surfaces will inspire future experimentation along this line.  In contrast, nonisometric origami can be probed using current synthesis techniques.  In this direction, we have shown that with simple building blocks, many complex shapes can be explored.  Thus taken together, we believe exploiting heterogeneity in nematic LCE and LCG sheets is a promising means of actuating complex shape, with many exciting avenues for further experimentation and possibly application.  

Paul Plucinsky is grateful for the support of the NASA Space Technology Research Fellowship.

\bibliographystyle{apsrev4-1}
\bibliography{plb} 

\end{document}